\documentstyle[12pt,epsf]{article}                                 

\begin{document}
\newcommand {\bea}{\begin{eqnarray}}
\newcommand {\eea}{\end{eqnarray}}
\newcommand {\be}{\begin{equation}}
\newcommand {\ee}{\end{equation}}

\title{
\begin{flushright}
\begin{small}
hep-th/9610077 \\
PUPT-1652 \\
UPR-719-T \\
\end{small}
\end{flushright}
\vspace{1.cm}
Extremal Branes as Elementary Particles
}

\author{Vijay Balasubramanian\\
\small Joseph Henry Laboratories\\
\small Princeton University\\
\small Princeton, NJ 08544 \\
\small e-mail: vijayb@puhep1.princeton.edu
\and 
Finn Larsen\\
\small Department of Physics and Astronomy\\
\small University of Pennsylvania\\
\small Philadelphia, PA 19104 \\
\small e-mail: larsen@cvetic.hep.upenn.edu
}

\date{}
\maketitle

\begin{abstract}
The supersymmetric p-branes of Type II string theory can be
interpreted after compactification as extremal black holes with zero
entropy and infinite temperature.  We show how the p-branes avoid this
apparent, catastrophic instability by developing an infinite mass gap.
Equivalently, these black holes behave like elementary particles: they
are dressed by effective potentials that prevent absorption of
impinging particles. In contrast, configurations with 2, 3, and 4
intersecting branes and their nonextremal extensions, behave
increasingly like conventional black holes.  These results extend and
clarify earlier work by Holzhey and Wilczek in the context of four
dimensional dilaton gravity.
\end{abstract}                                

\section{Introduction}

The recent interest in black holes has focussed on extremal
configurations with finite area and their non-extremal
generalizations because, in these cases, the finite entropy inferred 
from the area can be related to a microscopic counting in
string theory~\cite{sv1}(for reviews see~\cite{horowitz,juanthesis}).
Four dimensional black holes with finite area arise by 
compactification of configurations with at least 4 intersecting branes.
However, there are also supersymmetric configurations with
3, 2 and 1 intersecting branes. The corresponding black holes 
all have vanishing area and their formal temperatures, defined 
from the surface gravity, are 0, finite and infinite 
respectively~\cite{dyon}. The nonzero temperatures naively 
indicate that the compactified
supergravity solutions are semiclassically unstable since they would
radiate to produce naked singularities.  If we are to make
sense of compactified p-brane solutions there must therefore
be a mechanism that stabilizes these objects.

The analogous problem was confronted some time
ago in the context of dilaton gravity with action:
\begin{equation}
S = \int d^4x \sqrt{-g} \left(R - 2 (\nabla\phi)^2 + 
e^{-2a\phi} F^2 \right)
\label{eq:ac1}
\end{equation}
The extremal black hole solutions of this theory~\cite{ghs1,gm1} have
non-zero entropy for $a=0$ but vanishing entropy for $a>0$. Furthermore,
the formal temperature is zero for
$a<1$, finite for $a=1$ and infinite for $a>1$. The analogy with
intersecting branes is precise because classical solutions to
Eq.~\ref{eq:ac1} with $a=0,1/\sqrt{3},1,\sqrt{3}$ can be interpreted
in the context of type II string theory as marginally bound states of
elementary solutions with $a=\sqrt{3}$~\cite{dr1,dlr,dr2,bb1}.  These
extremal black holes with $a=\sqrt{3}$ played a crucial role in the duality
revolution~\cite{hull,witten95} and have since been interpreted at
weak coupling as D-branes~\cite{polch95a}.  Some of the required
marginal bound states have been shown to exist~\cite{vafa95a,sen95a}.

Extremal black holes with non-zero temperature inevitably develop
naked singularities and are therefore not physically acceptable.
Holzhey and Wilczek~\cite{hw} discovered that black holes with $a > 1$
have infinite mass-gaps - i.e., they support no finite energy
excitations! This causes the thermal description to break down and
moreover these black holes are unable to absorb any finite energy
impinging objects.  This implies that there is no radiation into these
modes either, via Kirchoff's Law.  In these senses, the $a>1$ holes
act like elementary particles, rather than as black holes.  For $a=1$
the mass gap is finite, leading to a situation where the black hole is
totally repulsive for objects below a particular critical energy.  This
also implies suppression of radiation at energies below this bound.
The finite temperature suggests  that these black holes might develop
naked singularities by radiating modes with energies higher than the
gap.   However, in~\cite{susy} it was argued that the thermodynamic
description breaks down close to extremality in such a way that this
is avoided.

In this paper, we will exhibit these phenomena directly in the context
of compactified p-brane solutions.  Holzhey and Wilczek found that
perturbations of the metric, dilaton and other fields displayed the
same qualitative behaviors as a minimally coupled spectator scalar; so
we limit ourselves to the latter case. The field equation governing
such a scalar is remarkably simple, despite a very general
background. Indeed, it reduces to the Schr\"{o}dinger equation for a
quantum particle in an attractive potential.  Excitation of the black
hole corresponds to absorption by the potential and repulsion of
impinging particles below a critical energy implies a gap in the
excitation spectrum.  As discussed above, the gap also implies a
breakdown in the thermal description of the object.  In our approach the
crucial distinction between black holes with $a>1$ and $a<1$ arises as
a simple consequence of the well-known feature of ${1\over r^s}$
potentials that they capture particles if $s>2$ but not if $s<2$.
 
The paper is organized as follows.  In section~\ref{sec:potentials},
we write the equations of motion for a spectator scalar, minimally
coupled to the ten-dimensional Einstein metric. We first consider a
single extremal p-brane and then several intersecting ones. In
section~\ref{sec:cross} we toroidally compactify intersecting branes
to make four-dimensional black holes and exhibit the repulsive
properties (or mass gaps) of one and two intersecting branes and the
absorption by three or four intersecting branes. In
Section~\ref{sec:nonext} we discuss scattering from non-extremal
p-branes and the approach to extremality.  The general wave equation
acquires a particularly simple form.  Finally, in
Section~\ref{sec:10d}, we derive the temperatures of uncompactified
extremal p-branes in 10 dimensions and show that they too develop
mass-gaps of the order of their temperatures.

\section{Effective Potentials}
\label{sec:potentials}

\subsection{Extremal p-Branes}
The 10 dimensional form of the type II action is
\begin{equation}
S = \int d^{10}x \sqrt{-g_S}  \left[
e^{-2\phi} (R_S + 4 (\nabla\phi)^2 ) - \frac{2}{(p+2)!} F_{p+2}^2 \right]
\label{eq:action}
\end{equation}
where $F$ is an RR (p+2)-form field strength and the subscripts $S$
indicate that the string metric is being used. The classical theory 
features extended p-brane solutions that are sources for the field
strengths in the action~\cite{hs}.\footnote{See also~\cite{dkl,ghkm}
for the solutions of~\cite{hs} in the isotropic coordinates used here.}
Their extremal incarnations are:
\begin{eqnarray}
ds^2_S &=& D^{-1/2} \left( -dt^2 + dx^i dx_i \right) +
         D^{1/2}  \left( dy^2 + y^2 \,d\Omega^2_{8-p} \right) \label{eq:sol1}\\
e^{-2\phi} &=& D^{(p-3)/2} \\
F &=& \frac{Q}{y^{8-p}} D^{-2} \: dt \wedge dx_1 \wedge \cdots
dx_p \wedge dy \\
D &=&  1 + \left(\frac{2}{7-p}\right) \frac{Q}{y^{7-p}} 
  \equiv \left( 1 + \frac{Q_p}{y^{7-p}} \right)  
\label{eq:Ddef}
\end{eqnarray}
The index $i$ runs from $1$ to $p$ and spans the brane volume.  
In the Einstein metric, related to the the string metric
through $g_{S\mu\nu} = g_{E\mu\nu}e^{\phi/2}$~\cite{dkl}, the 
metric Eq.~\ref{eq:sol1} becomes
\begin{equation}
ds^2_E =  D^{(p-7)/8} \left(-dt^2 + dx^i dx_i\right) 
        + D^{(p+1)/8} \left(dy^2 + y^2 d\Omega^2_{8-p}\right)
\label{eq:sol2}
\end{equation}
Note that the transverse part is conformally flat, with conformal
factor $C^2= D^{p+1\over 8}$, in the isotropic coordinates employed
here.  Now consider a spectator scalar field that is minimally
coupled to the background Einstein metric of Eq.~\ref{eq:sol2}.  The
equation of motion \be D_\mu\partial^\mu \chi =
{1\over\sqrt{-g_E}}\partial_\mu
(\sqrt{-g_E}g_E^{\mu\nu}\partial_\nu\chi) = 0
\label{eq:eom1}
\ee
simplifies because the volume element
\be
\sqrt{-g_E} = \left[ D^{{p+1\over 8}(9-p)} D^{{p-7\over 8}(p+1)}
\right]^{1\over 2} = D^{p+1\over 8}
\ee
is identical to the conformal factor of the transverse space 
$C^2=g_{yy}$.
Eq.~\ref{eq:eom1} acquires the simple form:
\begin{equation}
\left[ D \, \Box_{p+1} + \triangle_{9-p} \right] \chi = 0
\label{eq:eom}
\end{equation}
where $\Box_{p+1}$ is the Klein-Gordon operator (with Lorentzian 
signature) in the $p+1$ dimensional world volume theory
and $\triangle_{9-p}$ is the Laplacian (with Euclidian
signature) in the directions orthogonal to the brane.
The crucial simplification is that both are operators in {\it flat} space.

Introducing the spatial coordinates $\vec{x}$ and $\vec{y}$ that are tangent 
and transverse to the brane, respectively, and partially Fourier
transforming as 
$\chi= \exp{i(\omega t \pm \vec{k} \cdot \vec{x})} \psi(\vec{y})$
the expression becomes
\begin{equation}
\left( -\triangle_{9-p} - \frac{(\omega^2 - \vec{k}^2) Q_p}{y^{7-p}} 
\right) \psi = (\omega^2 - \vec{k}^2 ) \psi
\end{equation}
This is simply a Schr\"{o}dinger equation for a Coulomb problem in $9-p$
dimensions. There is therefore a wealth of results that can be drawn 
upon when analyzing the dynamics. However, before doing so, we will
show that analogous simplifications occur for more general 
brane configurations.

\subsection{Intersecting Extremal Branes}
The extremal branes can be viewed as building blocks that 
can combine into extremal, intersecting configurations that are 
also classical solutions of the theory in Eq.~\ref{eq:action}. 
This class of solutions includes
particularly interesting ones that can be interpreted as 
supersymmetric black holes with finite area after toroidal 
compactification~\cite{dyon,structure,cfthair1,cfthair2}.
Intersecting extremal branes can be constructed as follows. Extremality
is equivalent to preserving supersymmetry which in turn implies that 
p- and q-branes orthogonally intersecting on a k-brane must 
satisfy $p+q - 2k\equiv 0 \bmod 4 $~\cite{dnotes}.
In this case intersecting brane solutions are given by the 
``harmonic function rule''~\cite{harm,harm2}: multiply the harmonic functions
associated with individual p-branes in Eq.~\ref{eq:sol2} for
each metric component independently and similarly for the 
dilaton.  
For example, the intersection of a p-brane wrapped around the
$(1,\cdots,p)$ dimensions and another wrapped around the $(3,\cdots,p+2)$
dimensions gives rise to the fields
\begin{eqnarray}
ds_E^2 &=& 
\left(D_1 D_2\right)^{p-7\over 8} \left(-dt^2 + dx_3^2 + \cdots dx_p^2
\right) 
+ D_1^{p-7\over 8} D_2^{p+1\over 8} \left(dx_1^2 + dx_2^2\right)
\nonumber \\
&\mbox{}&+ D_1^{p+1\over 8} D_2^{p-7\over 8} 
\left(dx_{p+1}^2 + dx_{p+2}^2\right) 
+ \left(D_1 D_2\right)^{p+1\over 8} 
\left( dx_{p+3}^2 + \cdots dx_9^2 \right)
\label{eq:2psoln}
\\
e^{-2\phi}  &=& \left(D_1 D_2 \right)^{(p-3)/2}
\end{eqnarray}
where $D_1$ and $D_2$ are harmonic functions of the directions
transverse to both branes, {\it i.e.} $(p+3,\cdots,9)$. 
The corresponding expression in string metric is related
to Eq.~\ref{eq:2psoln} by the 
factor $e^{\phi/2} = \left(D_1 D_2 \right)^{-(p-3)/8}$.

In general, the Einstein metric for the dimensions parallel to the 
$i$th brane is multiplied by $D_i^{p-7\over 8}$ and the dimensions 
perpendicular to the brane are multiplied by $D_i^{p+8\over 8}$.  
The harmonic function is $D_i = 1 + Q_i/r^{s-2}$ 
where $s$ is the number of dimensions transverse to all the branes 
and $r$ is the coordinate radius in these dimensions. Finally,  
the expression for the dilaton is simply 
$e^{-2\phi} = \prod_i D_i^{(p_i-3)/2}$.

Consider a scalar field $\chi$ coupled minimally to the Einstein
metric of a general intersecting configuration. Again, the volume
factor
$\sqrt{-\det g_E}=\prod_i D_i^{(p_i + 1) /8} $
is identical to the conformal factor 
$C^2=\Pi_i D_i^{(p_i + 1) /8}$
of the space transverse to all the branes. Indeed, this is 
a consequence of the harmonic function rule and the 
corresponding result for individual p-branes. It follows that 
the Klein-Gordon equation reduces to 
$g_E^{\mu\nu} \partial_\mu\partial_\nu \chi = 0$ (in Cartesian 
coordinates). Multiplication by the conformal factor yields the 
equation of motion
\begin{equation}
\left[\sum_{i,j=0}^{9} H_i \, \eta^{ij} \, \partial_i \partial_j \right]
\chi = 0
\label{eq:geneom}
\end{equation}
where $H_i = \prod_{k_i} D_{k_i}$. The $k_i$ run over the indices
of the branes that are wrapped around the dimension $i$.
Specifically the $\partial_t^2$ term is multiplied by the product
of all the harmonics. For example, in the background of 
Eq.~\ref{eq:2psoln}, the expression Eq.~\ref{eq:geneom} becomes
\begin{equation}
\left[ D_1 D_2 \, \Box_{k+1}
+D_1 \left( \partial_1^2 + \partial_2^2 \right)
+D_2 \left(\partial_{p+1}^2 + \partial_{p+2}^2 \right)
+ \triangle \right] \chi = 0 
\end{equation}
where $\Box_{k+1}$ is the Klein-Gordon operator in the directions
parallel to both branes and $\triangle$  is the Laplacian in the
directions transverse to both branes.  Other instructive examples include 
the four-dimensional black holes built from D-branes considered
in~\cite{vf} {\it e.g.} four 3-branes wrapped around the
(123)(345)(146)(256) dimensions of a six-torus.  In this case 
the equation of motion Eq.~\ref{eq:geneom} becomes
\bea
[ -D_1 D_2 D_3 D_4 \partial_t^2 &+&
D_1 D_2 \partial_3^2 + D_1 D_3 \partial_1^2 + \nonumber \\
&+& D_1 D_4 \partial_2^2 + D_2 D_3 \partial_4^2 + 
D_2 D_4 \partial_5^2 + D_3 D_4 \partial_6^2 + \triangle ] \chi = 0
\eea
where $\triangle$ is the Laplacian in three dimensions.

The harmonic functions only depend on the transverse radius $r$; so
we can Fourier transform $\chi$ in the all non-transverse directions.
For intersecting branes the effective problem becomes scattering 
off a potential that includes several different powers of $1/r$, 
instead of a simple Coulomb problem. 

\section{Reflection and Absorption}
\label{sec:cross}
In this section we will discuss the scattering of neutral massless
scalars from configurations of up to four intersecting branes by
solving the wave equations derived in the preceding section.  We want
to study whether the branes can absorb impinging particles because, as
discussed in the introduction, inability to absorb a mode implies
inability to Hawking radiate into that mode also. To facilitate
comparison with the results of Holzhey and Wilczek~\cite{hw} we
toroidally compactify the intersecting branes to four dimensions.
This simply has the effect of replacing the harmonic functions $D = (1
+ Q_p/r^{s-2})$ from the previous section with $D = (1 + Q_p/r)$ where
$r$ is the radial distance in the non-compact dimensions (a numerical 
factor is absorbed in the definition of $Q_p$ ).  We will
consider scattering of scalar fields that are neutral under the
Kaluza-Klein U(1) gauge fields - i.e., fields that are independent of
the compactified coordinates.  For such neutral fields, we expand in
partial waves as $\chi= R_{l\omega} \, Y_{lm} \, e^{-i\omega t}$, and
find:
\begin{equation}
\left[
{\partial \over\partial r^2} + {2\over r}  
{\partial \over \partial r} - V_{\rm eff} 
\right]
R_{l\omega} =0~~;~~~ V_{\rm
eff}= -D_1 D_2 D_3  D_4 \omega^2 + {l(l+1)\over r^2}:
\label{eq:Veff}
\end{equation}
The functions $D_i$ have the harmonic form $1+{q_i\over r}$ with 
$s$ non-vanishing $q_i$'s in the case of $s$ intersecting branes.

\subsection{One Brane}
\label{sec:s1}
For a single brane the effective potential is:
\begin{equation}
V_{\rm eff} = -\omega^2 + { l(l+1) \over r^2} - {\omega^2 Q_1 \over r}
\label{eq:coul}
\end{equation}
Then the wave equation in Eq.~\ref{eq:Veff}
is formally identical to the Schr\"{o}dinger equation for a particle of
energy $E= \omega^2$ in an attractive Coulomb potential of charge
$\omega^2 Q_1$.  The exact solution to this problem is known, of course.
It is~\cite{landauqm}:
\begin{equation}
R_{l\omega} = {C_{\omega l} \over \Gamma(2l + 1)}
(2\omega r)^{l} e^{-i \omega r} \, 
F({i\omega Q_1\over 2} +l+1, 2l+2, 2i \omega r)
\label{eq:soln}
\end{equation}
where $F$ is the confluent hypergeometric function and $C_{\omega l}$ is
a normalization factor chosen so that $\int_0^\infty dr \, r^2 R_{l \omega'}
R_{l\omega} = 2\pi \delta(\omega - \omega')$.
The asymptotic form for large $r$ is:
\begin{equation}
R_{l \omega} \rightarrow {2\over r} \sin\left( 2\omega r + 
{Q_1\omega\over 2} \log{\omega r} - {l\pi \over 2} + \delta_l
\right)~~~~~;~~~~~
\delta_l = \arg  \Gamma(l+1 - i{\omega Q_1\over 2})
\label{eq:asymp}
\end{equation}
The relative phase shift between incoming and outgoing waves is:
\begin{equation}
2\Delta_l = -l\pi + 2\delta_l
\end{equation}
Since $\Delta_l$ is real, the incoming and outgoing flux at infinity
are equal and we can conclude that there is no absorption.  In other
words, a single extremal  p-brane compactified to four dimensions on a
6-torus is unable to absorb impinging particles!
The argument summarized here simply formalizes the well-known fact that 
Coulomb potentials have no absorptive part, even when they are attractive.

\subsection{Two Branes}
\label{sec:s2}
For two intersecting branes compactified on a six torus,
Eq.~\ref{eq:Veff} gives the effective potential:
\begin{equation}
V_{\rm eff} = -\omega^2 + { l(l+1)-\omega^2 Q_1 Q_2 \over r^2}
- {\omega^2 (Q_1+ Q_2) \over r}
\end{equation}
This is formally identical to the Coulomb potential Eq.~\ref{eq:coul}
with the effective angular momentum given through 
$L(L+1) = l(l+1)  - \omega^2 Q_1 Q_2$ and the Coulomb constant
modified according to $Q_1\rightarrow Q_1+Q_2$. It is still $l$ that
is quantized as a positive integer; so the
effective angular momentum $L$ is in general complex. An 
imaginary part develops for $Q_1Q_2 \omega^2 > (l+1/2)^2$.
Nevertheless the solution is still given by Eq.~\ref{eq:soln} 
with suitable replacements in the argument of the hypergeometric
function. From the asymptotic expansion Eq.~\ref{eq:asymp} which
remains valid for complex  of $l$ we find
the relative phase shift between incoming and outgoing waves scattering:
\begin{equation}
2\Delta_l = -L\pi + 2\delta_L ~~~~~~~
\delta_L = \arg \Gamma(L+1 - i{\omega(Q_1+Q_2)\over 2})
\end{equation}
When $L$ has an imaginary part the phase shift is complex and
the incoming flux is not equal to the outgoing flux. This indicates
that two intersecting branes absorb impinging particles with frequency
$\omega$ when $\omega^2 > (l + 1/2)^2/Q_1 Q_2$.

\subsection{More Branes}
When there are three branes present the effective 
potential in Eq.~\ref{eq:Veff} acquires a term of the form $1/r^3$ and
when there are four there will also be a $1/r^4$ term.
The corresponding quantum problems can not be solved exactly
but an approximate analysis suffices to determine
the qualitative behavior~\cite{landauqm}. The result is that 
attractive potentials, behaving as $r^{-s}$ for small $r$, are 
absorptive for $s>2$ and completely elastic for $s<2$ while the 
marginal case with $s=2$ depends on the competition between the 
potential and angular momentum~\cite{landauqm}. This implies that 
both three and four intersecting branes absorb impinging particles 
regardless of energy.

Rather than repeating the rigorous quantum mechanical analysis we find 
it instructive to consider the quasi-classical regime of large angular
momentum $l\gg 1$. Here the wave functions are of the quasi-classical
(WKB) form: 
\begin{equation}
\chi_0 = {1\over\sqrt{p_r}}\exp
(i\int^r p_r) 
\end{equation}
where $p_r$ is a slowly varying function that satisfies $-p_r^2=
V_{\rm eff}$.  The process can then be interpreted as a classical
particle subject to the potential $V_{\rm eff}$.  It is clear that for
$s=3,4$, the attractive potential $r^{-s}$ completely dominates the
centrifugal barrier and absorption follows for all but the largest $l$
(corresponding to the classical particle completely missing the black
hole).  However, for $s=2$, the attractive potential and the
centrifugal barrier are of equal importance and the more detailed
consideration in Sec.~\ref{sec:s2} is necessary (although the
semiclassical analysis happen to give the correct result). In the
final case of $s=1$ (a single brane), the attractive potential is
simply Coulombic.  Therefore the centrifugal barrier dominates for
large angular momentum, there is a classical turning point, and the
impinging particle is completely reflected.

The intuition deriving from the semiclassical approximation apparently
fails for the attractive Coulomb problem in the S-wave because here 
there is no angular momentum barrier and nevertheless the
potential reflects, as we saw in the exact treatment in 
Sec.~\ref{sec:s1}.
However, by the uncertainty relation, the kinetic energy
operator $p_r^2$ is bounded below by a term of order $({1\over
2r})^2$. So quantum uncertainty acts qualitatively as a classical
centrifugal barrier that can be overcome only by potentials that
diverges more rapidly than $1/r^2$ at the center or as $1/r^2$ with a
sufficiently large coefficient. This reconciles the intuitions of this 
section  with the  rigorous results presented above.

\subsection{Comparison With Results In Dilaton Gravity}
The results we have derived here can be compared with those of Holzhey
and Wilczek for dilaton black holes indexed by the parameter $a$.
They found absorption impossible for $a > 1$ and certain for $a < 1$
while in the case of $a=1$ the evidence was inconclusive.  However, as
described in the introduction, extremal black holes with parameters
$a=\sqrt{3}, 1 ,{1\over\sqrt{3}},0$ are identical to the four
dimensional manifestations of 1, 2, 3, and 4 intersecting extremal
branes respectively~\cite{dr1,dr2,bb1}. Our calculation is therefore
in perfect harmony with the Holzhey-Wilczek analysis~\cite{hw}.  We
find it very satisfying that the marginal case $a=1$ directly
corresponds to the more familiar marginality of $r^{-2}$ potentials.
It should be noted that gravity has repulsive properties in some
contexts, notably in the neighborhood of domain
walls~\cite{soleng}. We should therefore emphasize that the effective
potential Eq.~\ref{eq:Veff} is always attractive in the S-wave, even
for a single brane.  As shown in Sec~\ref{sec:s1}, absorption
by a single compactified brane is prevented by the long range nature
of the effective Coulomb-like interaction that governs the radial
motion of scalar fields in the p-brane metric, rather than by a
repulsive force.

\subsection{Hawking Radiation From Compactified Branes}
The lack of absorption of low-energy modes has consequences for
Hawking radiation. Indeed, for $s=1$ (a single brane) all finite
energy modes are reflected.  Now elementary thermodynamics implies
that the brane will not radiate into these modes either!  In the
marginal $s=2$ case (two branes), the precise condition for absorption
is $Q_1 Q_2\omega^2 > (l+{1\over 2})^2$, {\it i.e.}  such black holes
exhibit a finite mass gap. Naively Hawking radiation should be
perfectly thermal with finite temperature $T$ where
$T^{-1}=\beta=4\pi\sqrt{Q_1 Q_2}$ but the greybody factor implied by
the gap suppresses the emission amplitude completely for
$\beta\omega<2\pi(2l+1)$.  Modes of higher energy can be emitted
within this analysis, albeit with exponentially suppressed amplitudes.
We may fear that even the smallest amount of neutral radiation
inevitably exposes a naked singularity and that two intersecting
branes are therefore unstable despite the presence of a gap cutting
off low energy radiation.  However, the emission of a single high
energy mode is sufficient to change the Hawking temperature
substantially; so the thermal description is invalid in this regime
and a catastrophic fate is probably avoided~\cite{susy,multi}. Let us
also recall that the extremal black holes with $a=1$ have a
particularly simple description in weakly coupled string theory: they
are dual to elementary strings with the right movers in their ground
states~\cite{sen95}. These string states are protected by $BPS$
saturation and are absolutely stable. Indeed, there are no states in
string theory with the same charge but a lower mass.  At the present
level of treatment the quasi-classical approximation to Hawking
radiation does not capture this microscopic picture.

\section{Non-extremality}
\label{sec:nonext}
The results from the previous sections can be generalized to 
non-extremal black holes. Effective potentials are of comparable 
simplicity and facilitate an investigation of the
approach to extremality.   This enables us to discuss how a single
non-extremal compactified brane that is able to absorb particles
develops an infinite barrier in the extremal limit.
 
\subsection{Effective Potential}
Non-extremal versions of any of the extremal configurations of branes
from the previous sections are obtained by the introduction of yet
another harmonic function $f = (1 - \mu/r^{s-2})$ where $s$ is the
number of dimensions transverse to all the branes.  The non-extremal
metric is  modified  compared to the extremal case by the
substitutions~\cite{tseytlin96,pope96}: 
\begin{equation}
 g_{tt}\rightarrow fg_{tt}~~,~~~
g_{rr}\rightarrow f^{-1}g_{rr} 
\end{equation}
This preserves the volume element $\sqrt{-g_E}$ but the geometry of
the space transverse to all the branes changes non-trivially; so the
scalar wave equation does not immediately simplify as in previous
sections.  Note that in the non-extremal case the parameters $q_i$ of
the harmonic functions are non-trivially related to the physical
charge as $Q_i^2=q_i ( \mu+q_i )$.

Upon toroidal compactification to 4 dimensions the functions $f$
become $f = (1 - \mu/r)$ and the the metric exhibits a horizon at
$r_{+}=\mu$.   The minimally coupled scalar wave equation is:
\begin{equation}
[(-D_1 D_2 D_3 D_4 \partial_t^2 + 
f~{1\over r^2}
({1\over\sin\theta}\partial_\theta \sin\theta \partial_\theta+
\partial_\phi^2)
+{1\over r^2}f\partial_r (r^2 f\partial_r ) ]\chi =0
\end{equation}
For brevity we ignored possible dependence of $\chi$ on the compact
dimensions but generality could easily be restored. The {\it ansatz}
$\chi={1\over R}Y_{lm}\chi_0~e^{-i\omega t}$ where $R=rf^{1\over 2}$ 
yields:
\begin{equation}
(f^{-2}D_1 D_2 D_3 D_4 \omega^2 - f^{-1}{l(l+1)\over r^2}-
{1\over R}R^{\prime\prime})\chi_0 + \chi_0^{\prime\prime} =0
\end{equation}
where a prime denotes differentiation with respect to $r$.
Inserting the explicit expression for $R$ we find the effective
potential:
\begin{equation}
V_{\rm eff} = -f^{-2} D_1 D_2 D_3 D_4 \omega^2 + f^{-1}{l(l+1)\over r^2}
-{1\over 4}f^{-2}{\mu^2\over r^4}
\label{eqn:general}
\end{equation}
It is instructive to compare this expression with the extremal
potential Eq.\ref{eq:Veff}. The attractive potential towards the brane 
is stronger by the factor $f^{-2}$. This facilitates fall into the
black hole because the centrifugal barrier is only higher by the 
factor $f^{-1}$. There is also an additional attractive potential
that only depends on the non-extremality parameter $\mu$.

\subsection{The Approach to Extremality}
We now consider a single compactified non-extremal brane with finite
temperature and study how the gap in its radiation spectrum
switches off the radiation as the extremal limit is attained.  We
study the S-wave because higher angular momentum radiation is always
suppressed relative to the S-wave.  Introducing the shifted coordinate
$\rho=r-\mu$ that vanishes at the horizon and expanding in the
non-extremality parameter $\mu$, the s-wave potential becomes:
\begin{equation}
V_{\rm eff}^{\rm s-wave}= -\omega^2 (1+{q\over\rho})
-{1\over\rho^2}\omega^2(q+2\rho)\mu + {\cal O}(\mu^2)
\end{equation}
This effective potential contains attractive $1/\rho^2$ and $1/\rho$
pieces.  As shown in Sec.~\ref{sec:s2}, such a potential reflects all
modes with frequency $\omega$ where $\omega^2 q\mu< {1\over 4}$.
Appoaching extremality $\mu\rightarrow 0$ we conclude that
perturbations with any finite frequency $\omega$ reflect with
certainty! In this precise sense, the brane
exhibits an infinite mass gap in the extremal limit and the radiation
from the brane is completely supressed.

A single nonextremal brane is endowed with a finite temperature $T$,
where $T^{-1}=\beta = 4\pi\mu (1+{q\over\mu})^{1\over 2}$.  In terms
of the temperature, the condition for reflection from the non-extremal
brane is $\beta\omega< 2\pi$. This means that for any given finite $\mu$
there  could be radiation at very high energy energies above this
bound.  As the extremal limit is approached, the formal temperature diverges 
and the only modes that can be radiated have diverging energy.   
These modes have energies that are above the cutoff used to define 
the semiclassical theory; so radiation into modes reliably
described by the Hawking calculation will be completely suppressed
in the extremal limit.

\section{Scattering From Branes in 10 Dimensions}
\label{sec:10d}
In previous sections we studied branes compactified to four
dimensions. However, we expect that a similar analysis applies in more
general situation; so we proceed to consider uncompactified extremal
p-brane solutions.  In $10$ dimensions $6-$, $5-$, and lower branes
exhibit infinite, finite, and vanishing temperatures,
respectively~\cite{pope96}. This leads to an expectation that the
corresponding radial effective potentials behave like $r^{-1}$,
$r^{-2}$, and $r^{-s}$ with $s>2$ at short distances. To verify this
consider minimally coupled scalars that are independent of the
directions parallel to the brane. Employing the s-wave {\it ansatz}
$\chi = y^{-{8-p\over 2}}e^{-i\omega t}\chi_0$ the wave equation
eq.~\ref{eq:eom} becomes
\begin{equation}
-\frac{\partial^2}{\partial y^2}\chi_0 -
\omega^2(1 + \frac{Q}{y^{7-p}}) + \frac{(8-p)(6-p)}{4y^2}
 \chi_0 = 0
\end{equation}
The first term in the potential is attractive, the second is
repulsive. For 6-branes we have an attractive 
Coulomb potential, as expected. For $5$-branes the effective
problem involves a $r^{-2}$ potential, again as expected. Note, however,
that  details differ from the two intersecting branes in $4$ dimensions
because of the additional, repulsive ${3Q\over 4y^2}$ potential.
Nevertheless, all frequencies below a certain critical one are 
reflected; so the qualitative features remain unchanged.  
Similarly, for lower branes, the attractive $y^{-(7-p)}$ potential 
clearly dominates at short distances and leads to absorption at all 
energies. In sum, we find the expected qualitative picture.
It appears that the existence of mass gaps protecting extremal objects
with finite  temperatures is quite a general phenomenon.      

\section{Conclusion}
Our results were obtained by studying the dynamics of a minimally
coupled spectator scalar.  How universal is the qualitative behavior
of the appearance of mass gaps?  Holzhey and Wilczek~\cite{hw} found
in the context of dilaton gravity that metric, dilaton and gauge field
fluctuations all had the same qualitative scattering behavior.  In our
case it is simple to verify that minimal coupling to either the four
dimensional Einstein or string metrics gives the same wave equations
as the ones we study.  Moreover, momentum in the internal dimensions
({\it i.e.} charge under the Kaluza-Klein gauge fields) clearly leaves
the leading behavior of the effective potential close to the origin
unchanged. The qualitative behavior is therefore unmodified, although
the precise scattering coefficients will certainly vary for these
fields. These examples lead us to believe that the qualitative
behavior of repulsion or absorption of low-energy modes by 1,2,3 or 4
intersecting branes is generic.

In recent months, there has been a series of surprising results
demonstrating that the classical geometry around near-extremal black
holes affects the spectrum of Hawking radiation precisely so that
properties of non-perturbative string theory are
encoded~\cite{mathur96a,maldacena96b} (for a review
see~\cite{mathur96b} ).  In this paper, we have tried to reconcile the
non-zero temperatures of some classical solutions with their
interpretation as stable combinations of D-branes.  To do this we have
focussed on a curiously strong form of cosmic censorship that not only
hides the singularity behind a horizon, but also erects barriers.
This renders some compactified p-branes unable to absorb impinging
particles of arbitrary finite energy while erecting high classical
barriers around other extremal black holes.

\vspace{0.2in}
{\bf Acknowledgments:} 
We would like to thank C. Callan, M. Cveti\v{c}, S. Mathur, and
F. Wilczek for comments.  V.B. thanks Richard Sharkey for use of his
computer and Internet account.  F.L. is supported in part by DOE grant
AC02-76-ERO-3071.  V.B. is supported in part by DOE grant
DE-FG02-91-ER40671.

\end{document}